\documentclass[conference]{IEEEtran}
\IEEEoverridecommandlockouts
\usepackage{cite}
\usepackage{amsmath,amssymb,amsfonts}
\usepackage{subfigure}
\usepackage{algorithmic}
\usepackage{graphicx}
\usepackage{textcomp}
\usepackage[table]{xcolor}
\usepackage{pifont}
\usepackage{makecell}
\usepackage{multirow}
\def\BibTeX{{\rm B\kern-.05em{\sc i\kern-.025em b}\kern-.08em
    T\kern-.1667em\lower.7ex\hbox{E}\kern-.125emX}}
\begin{document}

\title{Interference-Asymmetric UAV Remote Control Links:  Measurements and Performance Evaluation
 \thanks{This work is supported in part by the NSF awards CNS-1939334 and CNS-2332835.}
}

\author{Donggu Lee$^1$, Sung Joon Maeng$^2$, Ozgur Ozdemir$^1$, Mani Bharathi Pandian$^3$, and Ismail Guvenc$^1$ \\
$^1$Department of Electrical and Computer Engineering, North Carolina State University, Raleigh, NC, USA \\
$^2$Department of Electrical and Electronic Engineering, Hanyang University, Ansan, South Korea \\
$^3$Skydio, Inc., Redwood City, CA, USA \\
E-mail: \{dlee42, oozdemi, iguvenc\}@ncsu.edu, sjmaeng@hanyang.ac.kr, mani.pandian@skydio.com}

% \author{\IEEEauthorblockN{Donggu Lee}
% \IEEEauthorblockA{\textit{Department of Electrical and Computer Engineering} \\
% \textit{North Carolina State University}\\
% Raleigh, NC, USA \\
% dlee42@ncsu.edu}
% \and
% \IEEEauthorblockN{Sung Joon Maeng}
% \IEEEauthorblockA{\textit{Department of Electrical and Computer Engineering} \\
% \textit{North Carolina State University}\\
% Raleigh, NC, USA \\
% smaeng@ncsu.edu}
% \and
% \IEEEauthorblockN{Ismail Guvenc}
% \IEEEauthorblockA{\textit{Department of Electrical and Computer Engineering} \\
% \textit{North Carolina State University}\\
% Raleigh, NC, USA \\
% iguvenc@ncsu.edu}
% }

\maketitle

\begin{abstract}
% Command and control of uncrewed aerial vehicles (UAVs) is often realized through air-to-ground (A2G) remote control (RC) links that operate in industrial, scientific, and medical (ISM) bands. While wireless fidelity (Wi-Fi) technology is commonly used for UAV RC links, cellular systems such as long-term evolution (LTE) and 5G technologies have also been recently considered for the same purpose. 
Reliable and secure connectivity is crucial for remote control (RC) and uncrewed aerial vehicles (UAVs) links. A major problem for UAV RC links is that interference sources within the coverage may degrade the link quality. Such interference problems are a higher concern for the UAV than the RC unit on the ground due to the UAV being in line of sight (LoS) with a larger number of interference sources. As a result, lost hybrid automatic repeat request (HARQ) indicators (ACK/NACK) feedback in the uplink (UL, RC to UAV) may degrade the downlink (DL, UAV to RC) throughput. To get physical evidence for our interference asymmetry argument, we first conducted a measurement campaign using a helikite platform at the Main Campus area of NC State University during the 2024 Packapalooza festival. Subsequently, we evaluated the throughput impact of the loss of HARQ indicator feedback caused by UL asymmetry using MATLAB long-term-evolution (LTE) and fifth-generation (5G) toolboxes. Our numerical results confirm that UL interference asymmetry substantially degrades the throughput performance due to the loss of HARQ indicator feedback.
\end{abstract}

\begin{IEEEkeywords}
A2G networks, drone, measurement, remote control, throughput, UAV.
\end{IEEEkeywords} 

\section{Introduction}
Reliable and secure connectivity is essential for remote control (RC) and mission-critical uncrewed aerial vehicle (UAV)-based application scenarios. To achieve these, wireless fidelity (Wi-Fi)~\cite{7986413}, and recently, long-term-evolution (LTE)~\cite{8287894, DJI_paper} technologies are commonly adopted for air-to-ground (A2G) RC links. However, ensuring reliable and secure connectivity in dynamic aerial environments introduces additional challenges.

\begin{figure}[t!]
    \centering
    \includegraphics[width=0.99\columnwidth]{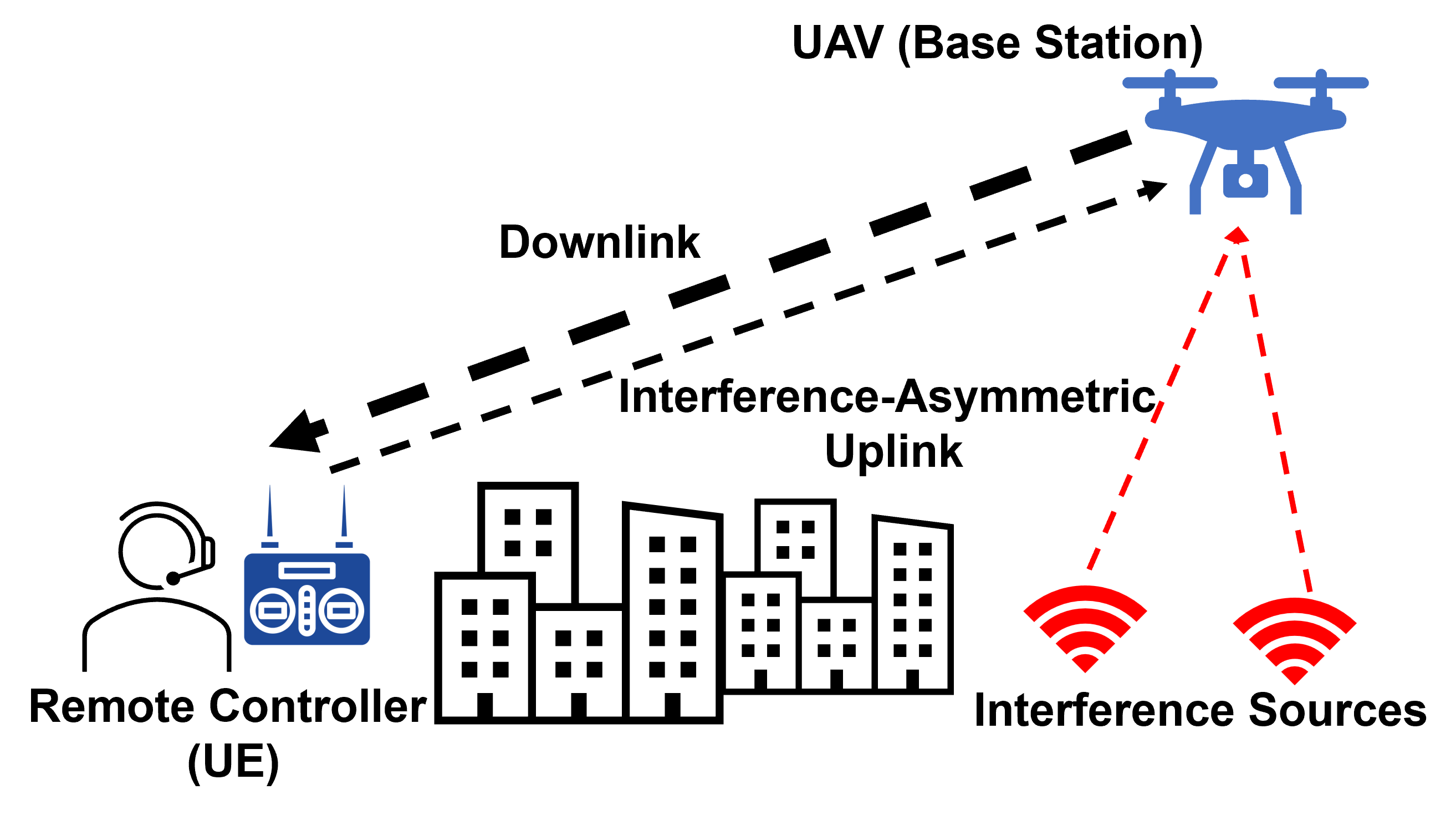}%\vspace{-0.3cm}
    \caption{A2G network model with interference-asymmetric UL. The UE is isolated more from interference sources than the UAV in the air. See the thinner arrow in the interference-asymmetric UL highlights the lower quality than the DL.}%\vspace{-0.5cm}
    \label{fig:A2G_model}
\end{figure}

A critical requirement for such air-to-ground (A2G) links is to maintain long-distance link connectivity between the RC on the ground and the UAV. Given this perspective, interference is an essential factor affecting the stability of A2G links. Unlike a general cellular base station on the ground, UAVs operating as aerial base stations are more vulnerable to line-of-sight (LoS) interference due to their elevated locations, as seen in Fig.~\ref{fig:A2G_model}. This brings a unique uplink (UL, RC to UAV) and downlink (DL, UAV to RC) asymmetry. While the RC unit on the ground has relatively robust channel conditions, the UAV in the air experiences an unfavorable condition caused by severe interference. As a result, the reliability of the hybrid automatic repeat request (HARQ) indicator (ACK/NACK) feedback through the UL is substantially degraded by the UL asymmetry compared to the conventional cellular networks.

In~\cite{PLE_table_asymmetry}, the asymmetric UL and DL in the aerial networks are investigated using measurements of the path loss coefficient. According to this study, the path loss coefficients over the UL and DL are derived as $2.51$ and $2.32$, respectively. These results highlight that a higher path loss coefficient in the UL can bring a mismatch between real-world and symmetric assumed scenarios. 

A similar study can be found in~\cite{asymmetry_UAV, asymmetry_uav_journal}. In~\cite{asymmetry_UAV}, the UL asymmetry of UAV-assisted wireless sensor networks is studied. The authors evaluate UL asymmetry using performance metrics of packet reception rate and received signal strength indicator (RSSI) to validate UL asymmetry, which can cause degradation of data collection efficiency and network stability. In~\cite{asymmetry_uav_journal}, the UL asymmetry is analyzed considering the mobility of the UAV. Due to a higher path loss coefficient in the UL, the mobility of UAVs intensifies the effect of dynamic path loss variation, leading to a wider range of fluctuation in UL channel conditions than in the DL.

As such, in this study, we focus on an UL interference-asymmetric scenario. To obtain physical evidence, we conduct a measurement campaign using a helikite platform in the Main Campus area at NC State University during the 2024 Packapalooza festival. We consider LTE, fifth-generation (5G), and industry, science, and medical (ISM) bands for our measurement campaign. Consequently, we evaluate LTE and 5G physical DL shared channel (PDSCH) throughput performance using MATLAB simulation, considering the loss of HARQ indicator feedback in the physical UL control channel (PUCCH) due to the UL asymmetry.

The rest of the paper is organized as follows. The setups and results of the helikite measurement campaign are demonstrated in Section~\ref{CH:measurement}. The system model and performance evaluation scenario for A2G UAV RC are described in Section~\ref{CH:sys}. Performance evaluation scenarios and numerical results are presented in Section~\ref{ch:numerical_analysis}. Finally, the last section concludes the paper.

\section{Helikite Measurement Campaign}\label{CH:measurement}
To obtain physical evidence of our UL asymmetry model, we conducted a measurement campaign using a helikite platform in an urban area~\cite{maeng_2024_Packapalooza, Maeng_2025}. 

\subsection{Helikite Measurement Campaign Setup}
\begin{figure}[t!]\vspace{0.05in}
    \centering
    \includegraphics[width=0.95\columnwidth]{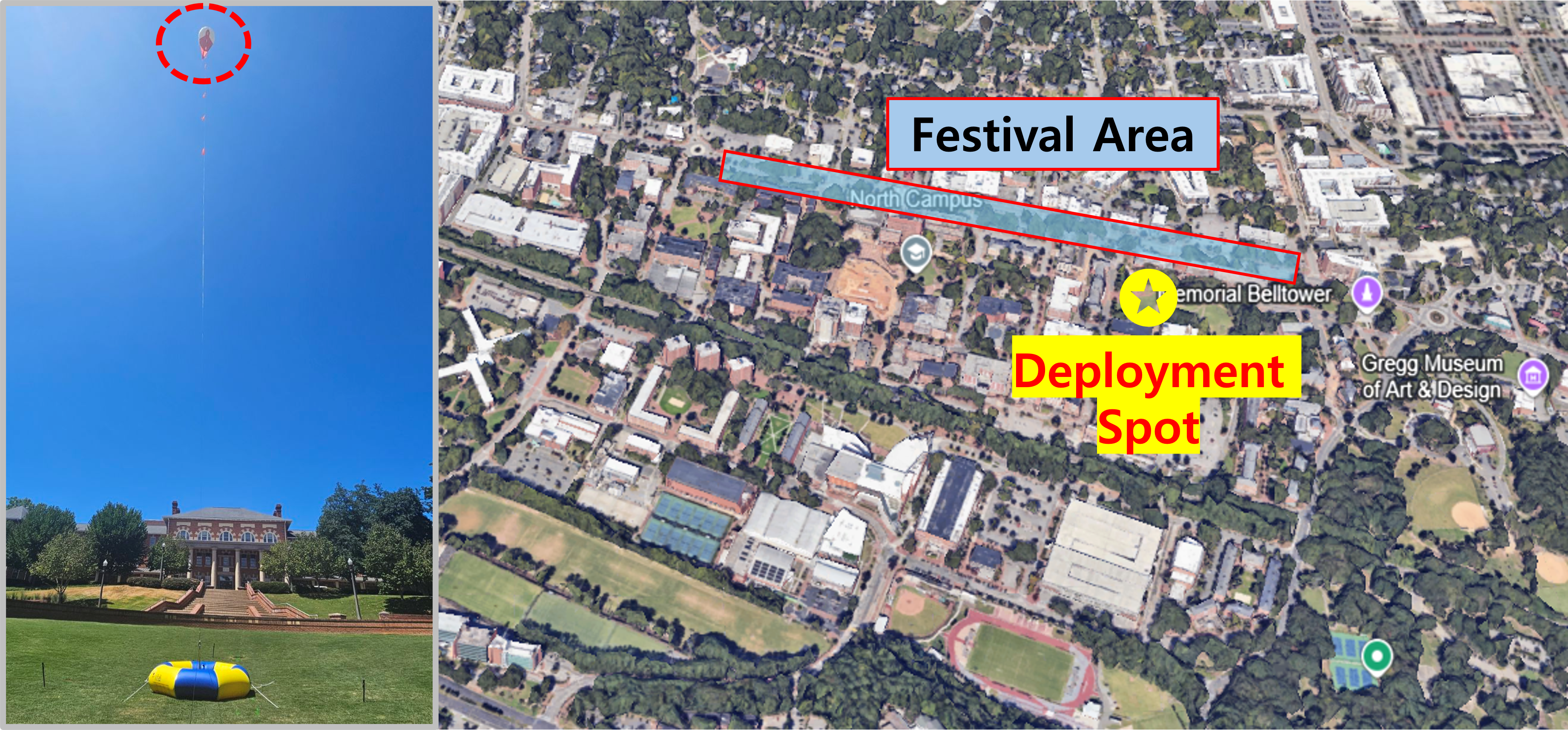}   %\vspace{-0.2cm}   
    \caption{Locations of deployment spot of helikite and measurement campaign area in 2024 Packapalooza Festival.}%\vspace{-0.3cm}
    \label{fig:measurement_campaign_pictures}
\end{figure}

The helikite measurement campaign was conducted at the Main Campus area of NC State University during the $2024$ Packapalooza festival (in August $2024$). The helikite deployment spot and the festival area for the measurement campaign are highlighted in Fig.~\ref{fig:measurement_campaign_pictures}. 

\begin{figure}[t!]
    \centering
    \subfigure[3D trajectory.]{\includegraphics[trim={0.6cm, 0, 1.3cm, 0.35cm},clip, width=0.47\columnwidth]{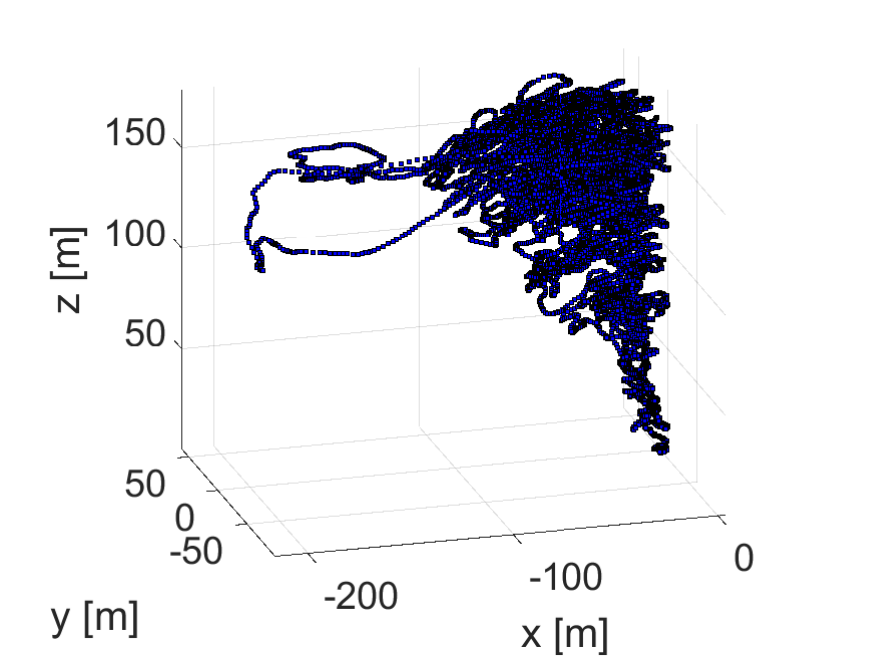}
    \label{fig:3D_trajectory_2024_packapalooza}}
    \subfigure[Altitude over time.]{\includegraphics[trim={0.2cm, 0, 1.3cm, 0.35cm},clip, width=0.47\columnwidth]{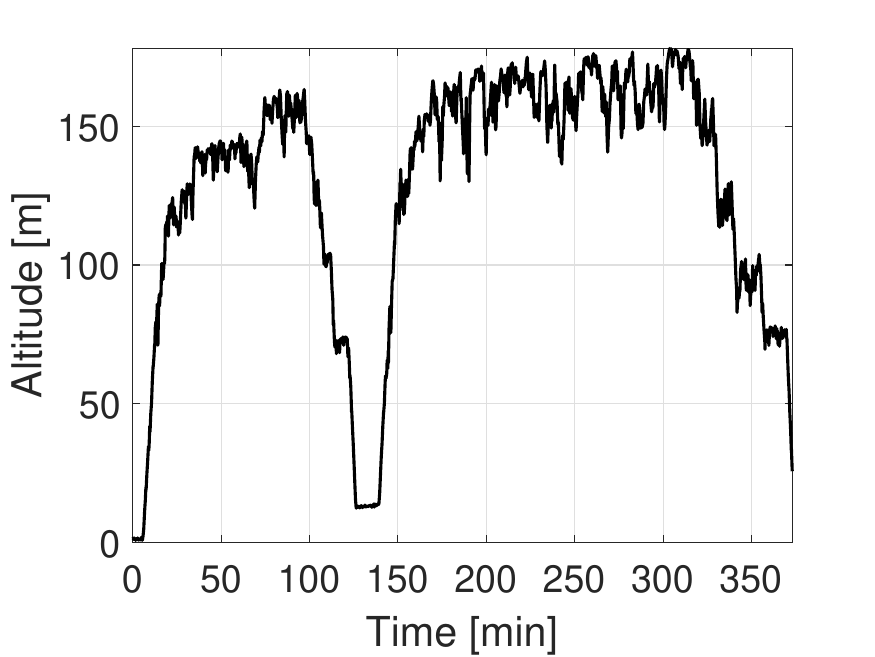}
    \label{fig:altitude_time_2024_packapalooza}}    %\vspace{-0.2cm}
    \caption{3D trajectory and altitude over time of helikite measurement campaign in the 2024 Packapalooza area.}%\vspace{-0.4cm}
    \label{fig:experiement_setups_helikite}
\end{figure}

\begin{figure*}[t!]
    \centering
    \subfigure[RSS versus altitude: Band $12$.]{\includegraphics[trim={0.1cm, 0, 1.3cm, 0.2cm},clip, width=0.29\textwidth]{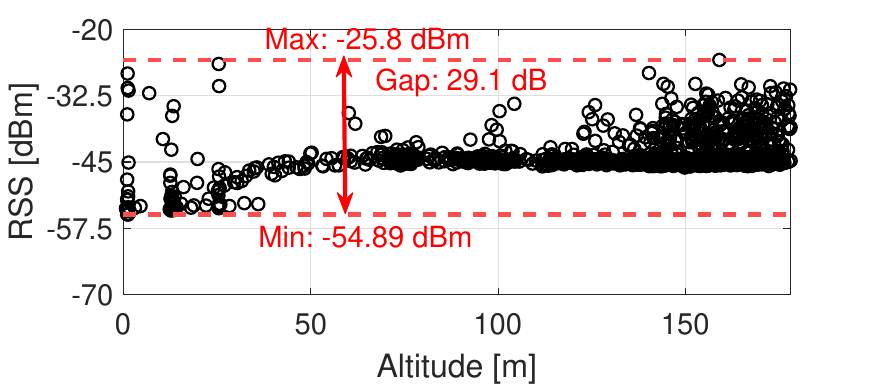}
    \label{fig:rss_altitude_2024_Packapalooza_LTE_band12}}
    \subfigure[RSS versus time with altitude: Band $12$.]{\includegraphics[trim={0.1cm, 0, 0cm, 0cm},clip, width=0.32\textwidth]{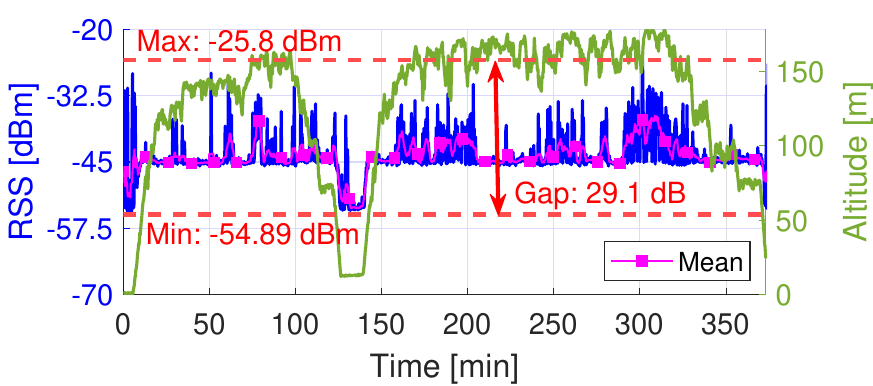}\label{fig:rss_time_altitude_2024_Packapalooza_LTE_band12}}
    \subfigure[RSS versus frequency: Band $12$.]{\includegraphics[trim={0.1cm, 0, 1.0cm, 0.2cm},clip, width=0.29\textwidth]{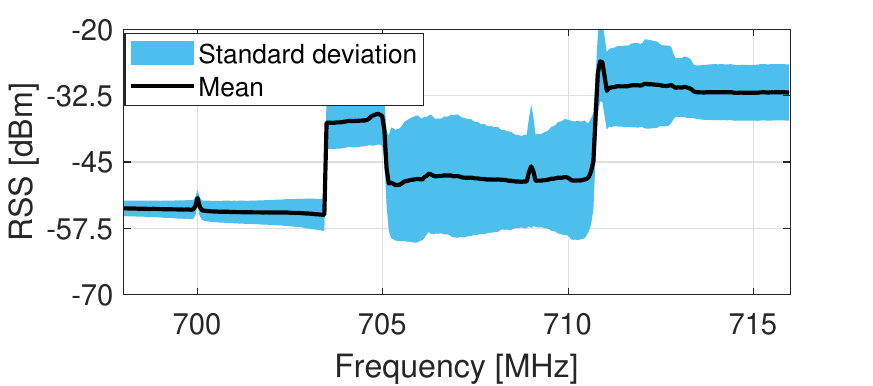}\label{fig:rss_frequency_2024_Packapalooza_LTE_band12}}
    
    \subfigure[RSS versus altitude: Band $13$.]{\includegraphics[trim={0.1cm, 0, 1.3cm, 0.2cm},clip, width=0.3\textwidth]{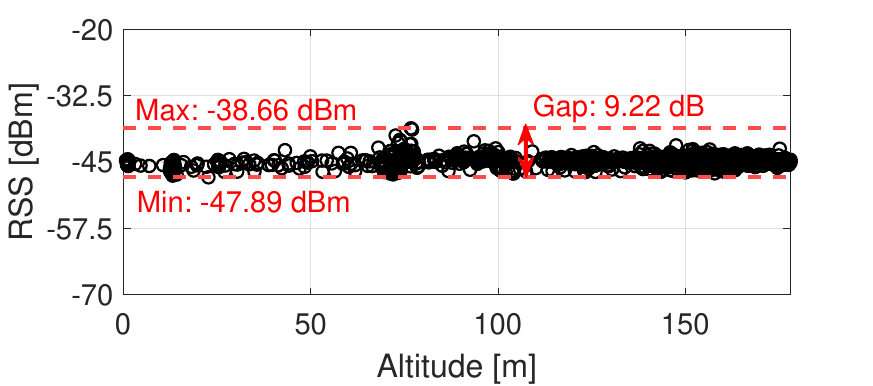}
    \label{fig:rss_altitude_2024_Packapalooza_LTE_band13}}
    \subfigure[RSS versus time with altitude: Band $13$.]{\includegraphics[trim={0.1cm, 0, 0cm, 0cm},clip, width=0.3\textwidth]{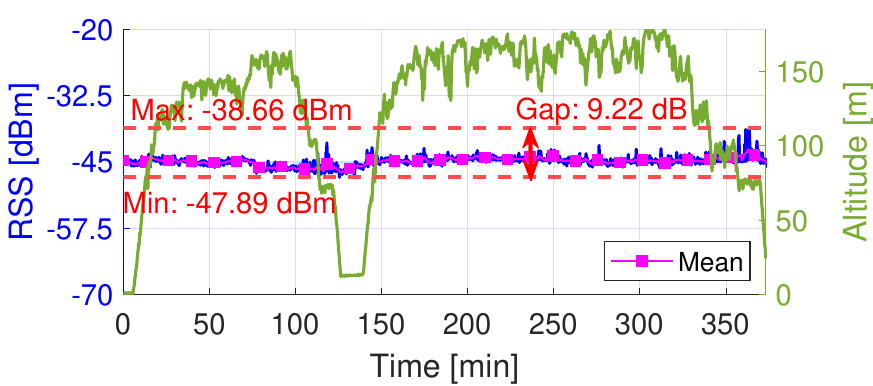}
    \label{fig:rss_time_altitude_2024_Packapalooza_LTE_band13}}
    \subfigure[RSS versus frequency: Band $13$.]{\includegraphics[trim={0.1cm, 0, 1.0cm, 0.2cm},clip, width=0.3\textwidth]{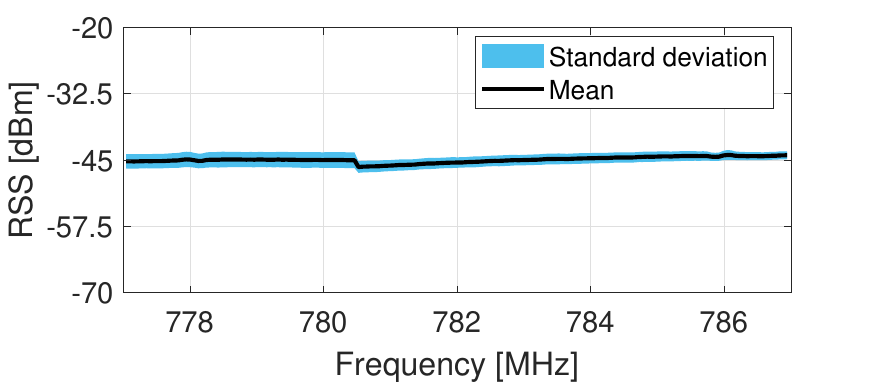}}
    
    \subfigure[RSS versus altitude: Band $14$.]{\includegraphics[trim={0.1cm, 0, 1.3cm, 0.2cm},clip, width=0.3\textwidth]{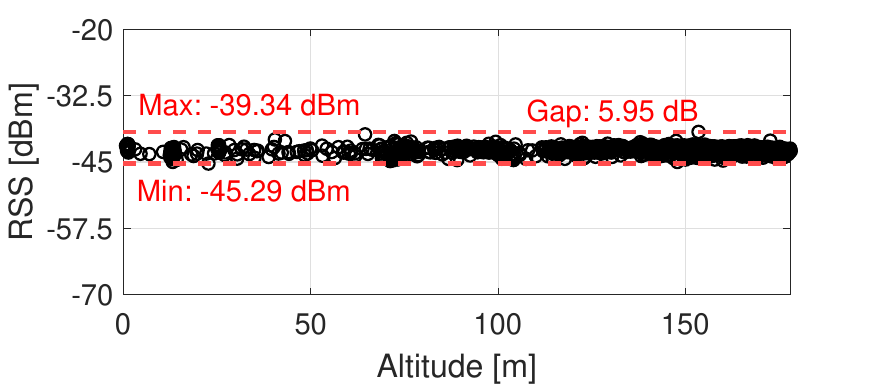}
    \label{fig:rss_altitude_2024_Packapalooza_LTE_band14}}
    \subfigure[RSS versus time with altitude: Band $14$.]{\includegraphics[trim={0.1cm, 0, 0cm, 0cm},clip, width=0.3\textwidth]{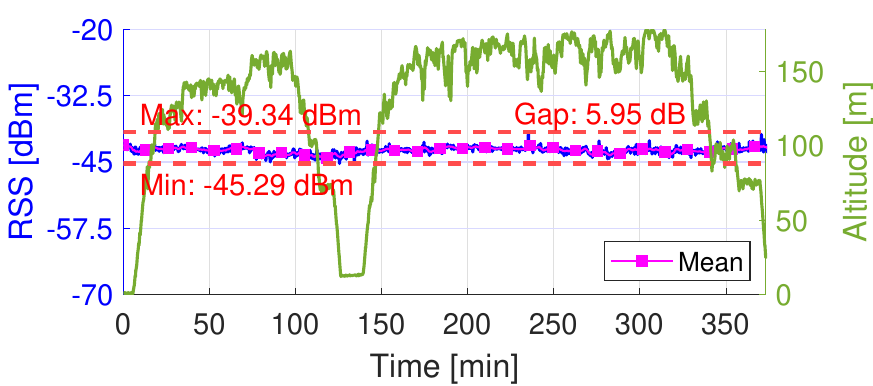}
    \label{fig:rss_time_altitude_2024_Packapalooza_LTE_band14}}
    \subfigure[RSS versus frequency: Band $14$.]{\includegraphics[trim={0.1cm, 0, 1.0cm, 0.2cm},clip, width=0.3\textwidth]{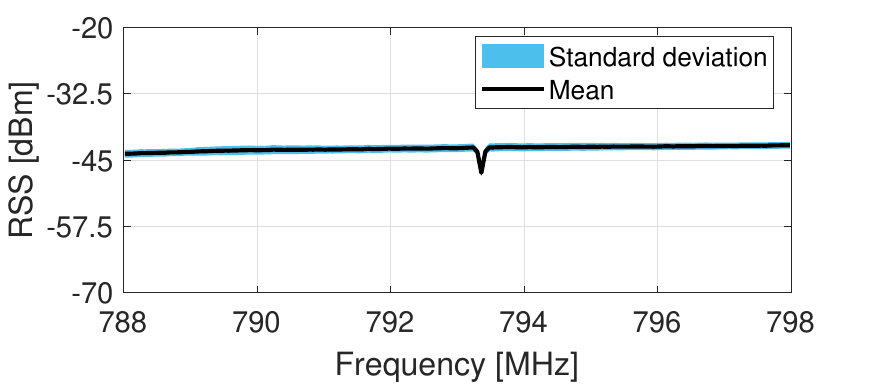}}
    
    \subfigure[RSS versus altitude: Band $41$.]{\includegraphics[trim={0.1cm, 0, 1.3cm, 0.2cm},clip, width=0.3\textwidth]{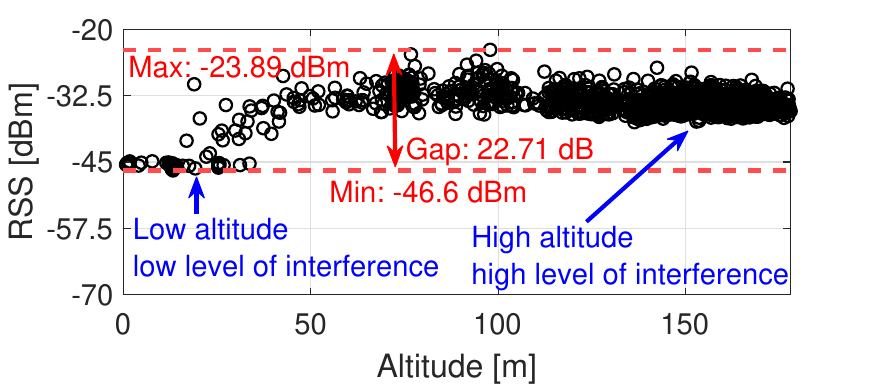}
    \label{fig:rss_altitude_2024_Packapalooza_LTE_band41}}
    \subfigure[RSS versus time with altitude: Band $41$.]{\includegraphics[trim={0.1cm, 0, 0cm, 0cm},clip, width=0.3\textwidth]{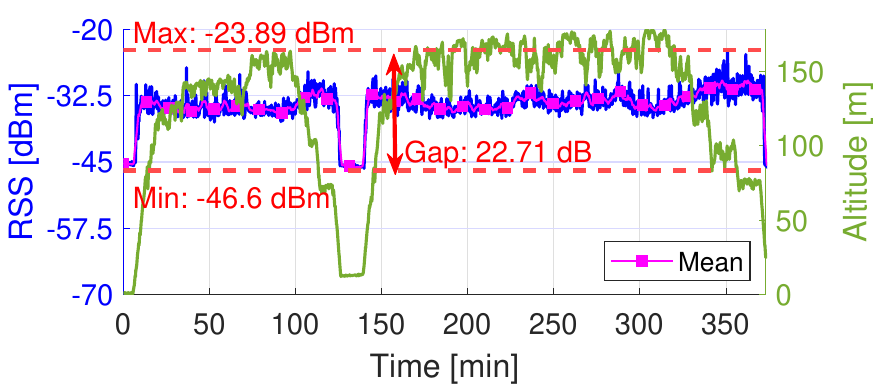}
    \label{fig:rss_time_altitude_2024_Packapalooza_LTE_band41}}
    \subfigure[RSS versus frequency: Band $41$.]{\includegraphics[trim={0.1cm, 0, 1.0cm, 0.2cm},clip, width=0.3\textwidth]{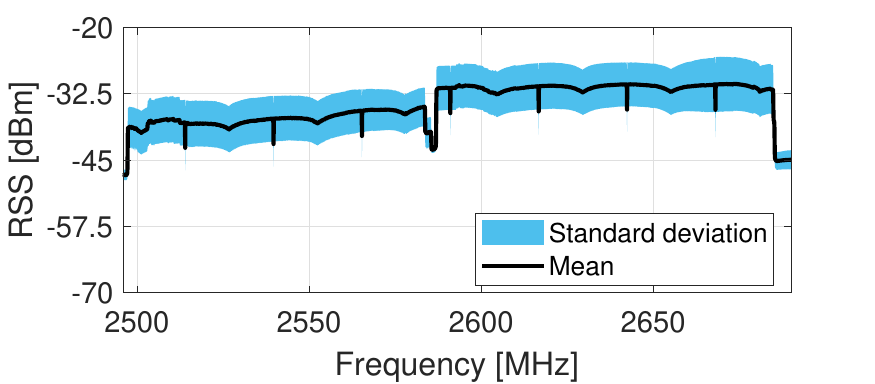}\label{fig:rss_frequency_2024_Packapalooza_LTE_band41}}%\vspace{-0.2cm}
    
    \caption{RSS measurements of LTE bands in the urban areas of the 2024 Packapalooza Festival.}%\vspace{-0.4cm}
    \label{fig:rss_altitude_freq_2024_Packapalooza_LTE}
\end{figure*}

\begin{figure*}[t!]
    \centering
    \subfigure[RSS versus altitude: Band n5.]{\includegraphics[trim={0.1cm, 0, 1.3cm, 0.2cm},clip, width=0.29\textwidth]{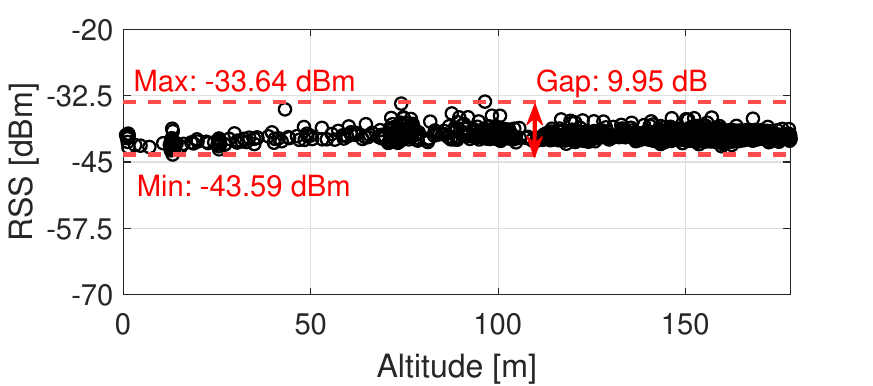}
    \label{fig:rss_altitude_2024_Packapalooza_5G_bandn5}}
    \subfigure[RSS versus time with altitude: Band n5.]{\includegraphics[trim={0.1cm, 0, 0cm, 0cm},clip, width=0.32\textwidth]{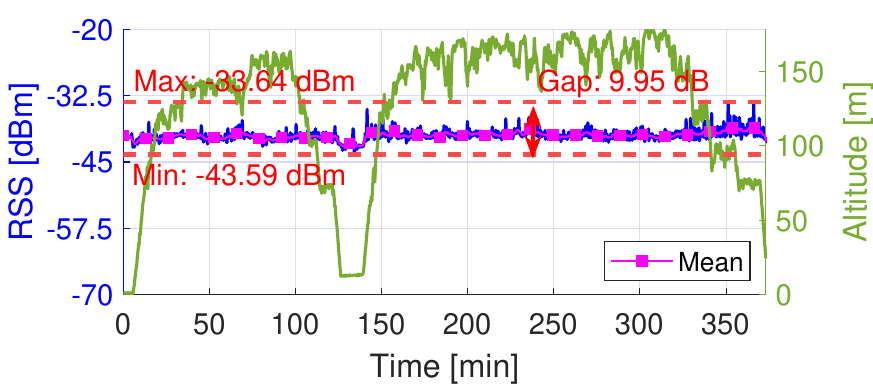}
    \label{fig:rss_time_altitude_2024_Packapalooza_5G_bandn5}}
    \subfigure[RSS versus frequency: Band n5.]{\includegraphics[trim={0.1cm, 0, 1.3cm, 0.2cm},clip, width=0.29\textwidth]{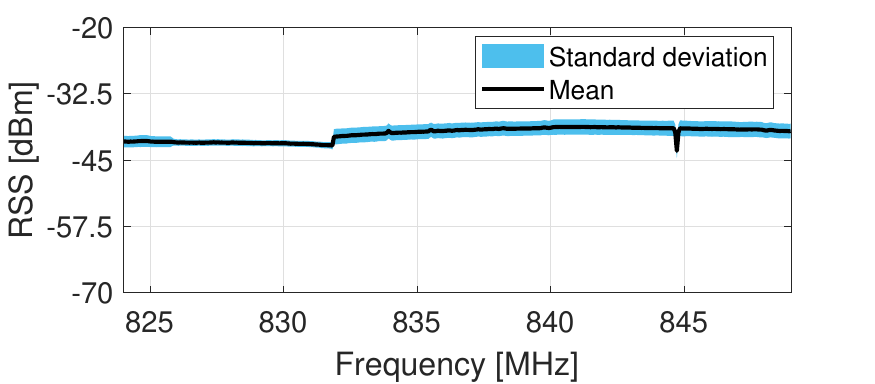}}
    
    \subfigure[RSS versus altitude: Band n71.]{\includegraphics[trim={0.1cm, 0, 1.3cm, 0.2cm},clip, width=0.29\textwidth]{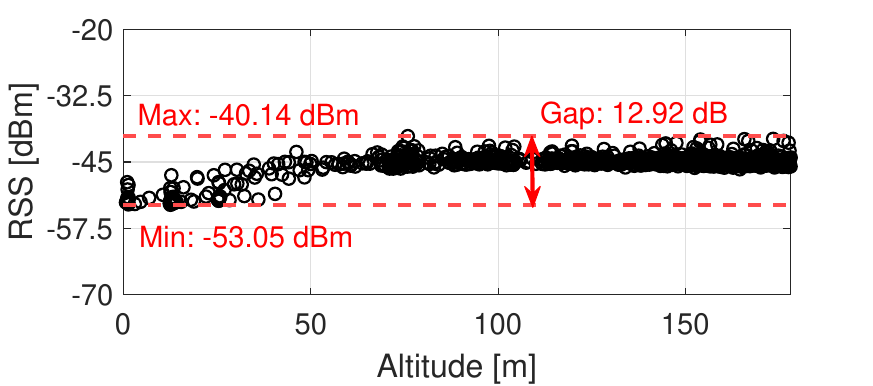}
    \label{fig:rss_altitude_2024_Packapalooza_5G_bandn71}}
    \subfigure[RSS versus time with altitude: Band n71.]{\includegraphics[trim={0.1cm, 0, 0cm, 0cm},clip, width=0.32\textwidth]{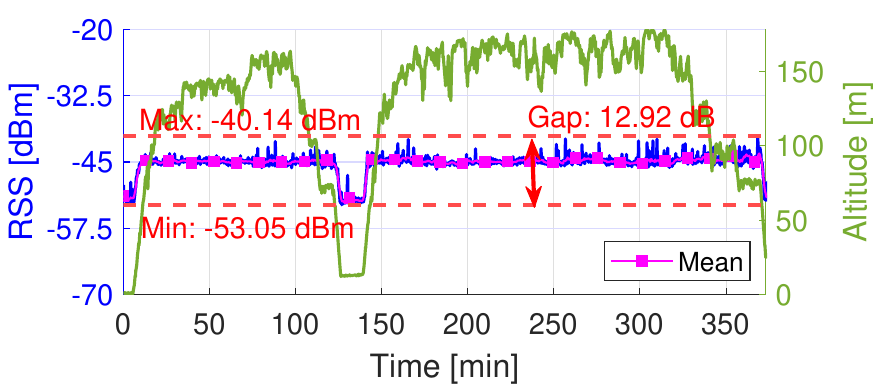}
    \label{fig:rss_time_altitude_2024_Packapalooza_5G_bandn71}}
    \subfigure[RSS versus frequency: Band n71.]{\includegraphics[trim={0.1cm, 0, 1.3cm, 0.2cm},clip, width=0.29\textwidth]{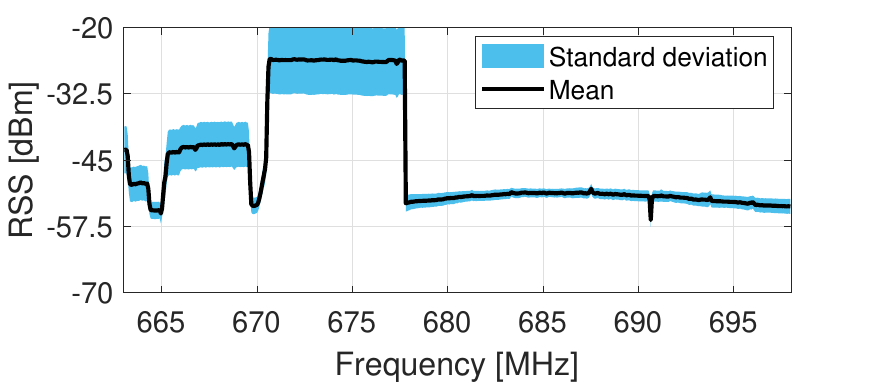}}
    
    \subfigure[RSS versus altitude: Band n77.]{\includegraphics[trim={0.1cm, 0, 1.3cm, 0.2cm},clip, width=0.29\textwidth]{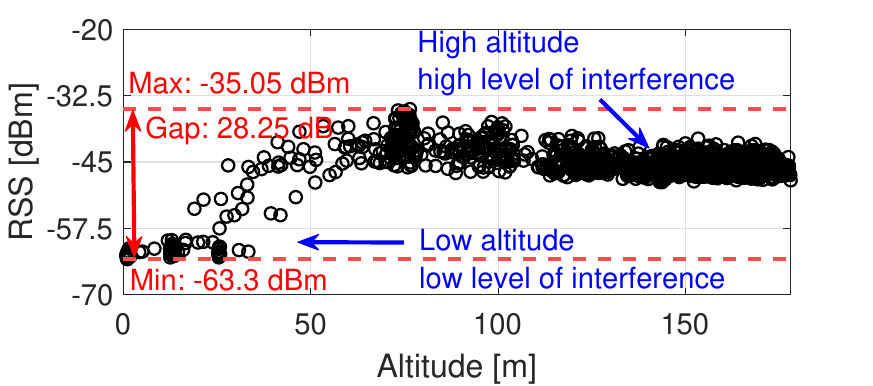}
    \label{fig:rss_altitude_2024_Packapalooza_5G_bandn77}}
    \subfigure[RSS versus time with altitude: Band n77.]{\includegraphics[trim={0.1cm, 0, 0cm, 0cm},clip, width=0.32\textwidth]{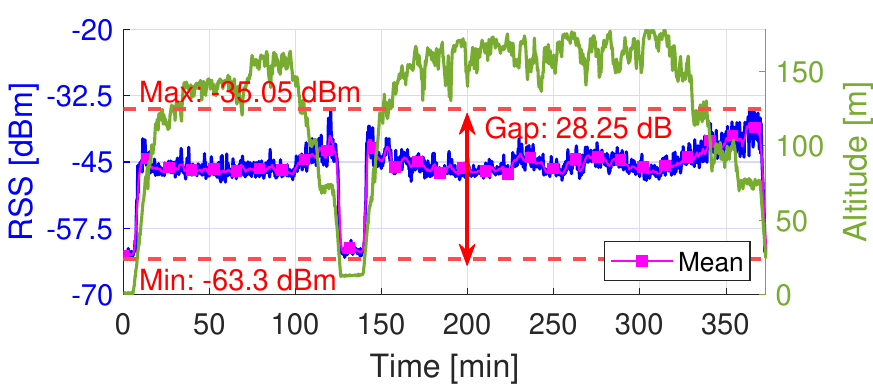}
    \label{fig:rss_time_altitude_2024_Packapalooza_5G_bandn77}}
    \subfigure[RSS versus frequency: Band n77.]{\includegraphics[trim={0.1cm, 0, 1.3cm, 0.2cm},clip, width=0.29\textwidth]{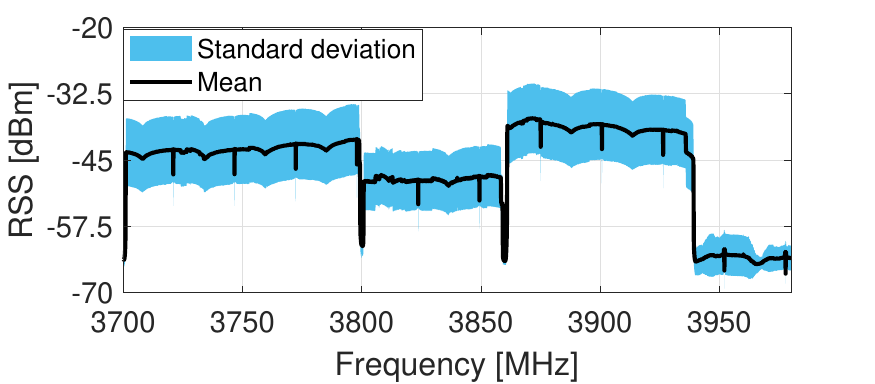}}%\vspace{-0.2cm}
    \caption{RSS measurements of 5G bands in the urban areas of the 2024 Packapalooza Festival.}%\vspace{-0.3cm}
    \label{fig:rss_altitude_freq_2024_Packapalooza_5G}
\end{figure*}

\begin{figure*}[t!]
    \centering
    \subfigure[RSS versus altitude: Channel 1.]{\includegraphics[trim={0.1cm, 0, 1.3cm, 0.3cm},clip, width=0.29\textwidth]{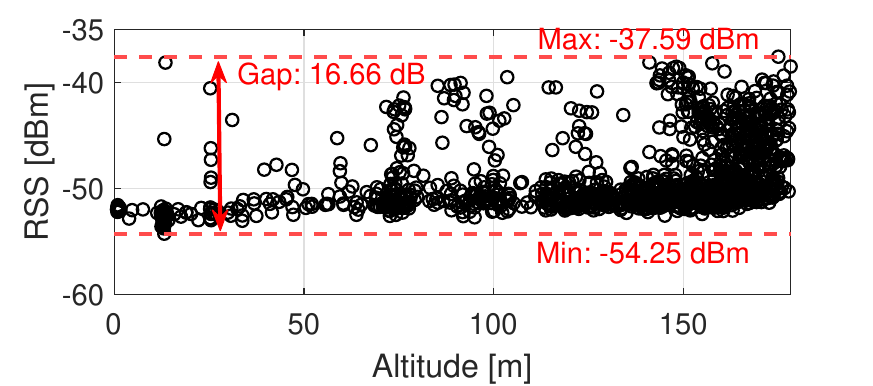}
    \label{fig:rss_altitude_2024_Packapalooza_WiFi_channel1}}
    \subfigure[RSS versus time with altitude: Channel 1.]{\includegraphics[trim={0.1cm, 0, 0cm, 0cm},clip, width=0.32\textwidth]{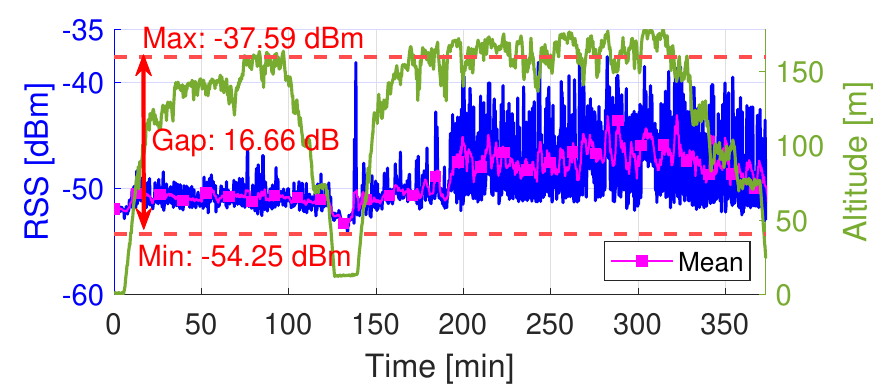}
    \label{fig:rss_time_altitude_2024_Packapalooza_WiFi_channel1}}
    \subfigure[RSS versus frequency: Channel 1.]{\includegraphics[trim={0.1cm, 0, 0.8cm, 0.3cm},clip, width=0.29\textwidth]{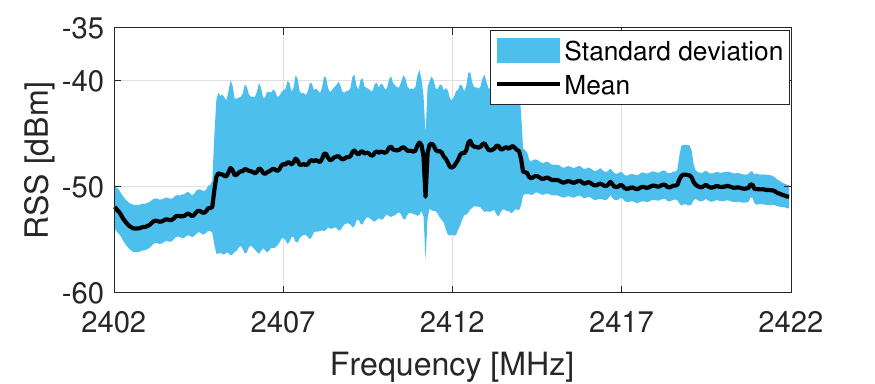}}
    
    \subfigure[RSS versus altitude: Channel 6.]{\includegraphics[trim={0.1cm, 0, 1.3cm, 0.3cm},clip, width=0.29\textwidth]{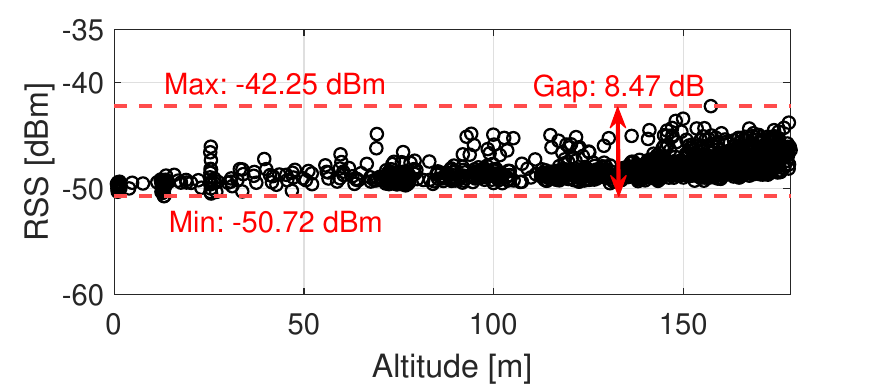}
    \label{fig:rss_altitude_2024_Packapalooza_WiFi_channel6}}
    \subfigure[RSS versus time with altitude: Channel 6.]{\includegraphics[trim={0.1cm, 0, 0cm, 0cm},clip, width=0.32\textwidth]{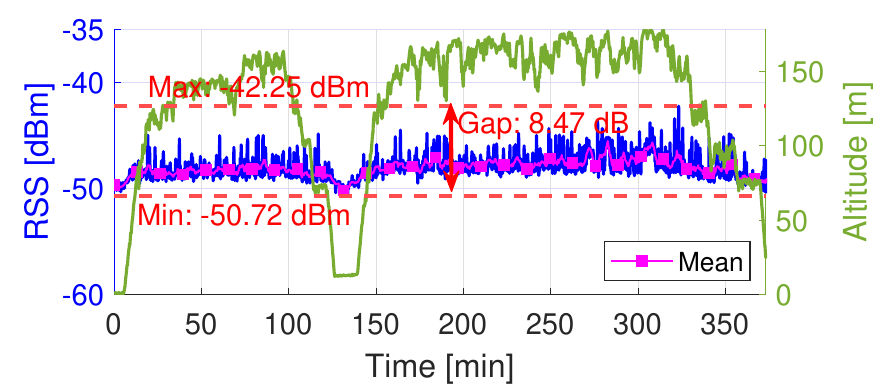}
    \label{fig:rss_time_altitude_2024_Packapalooza_WiFi_channel6}}
    \subfigure[RSS versus frequency: Channel 6.]{\includegraphics[trim={0.1cm, 0, 0.8cm, 0.3cm},clip, width=0.29\textwidth]{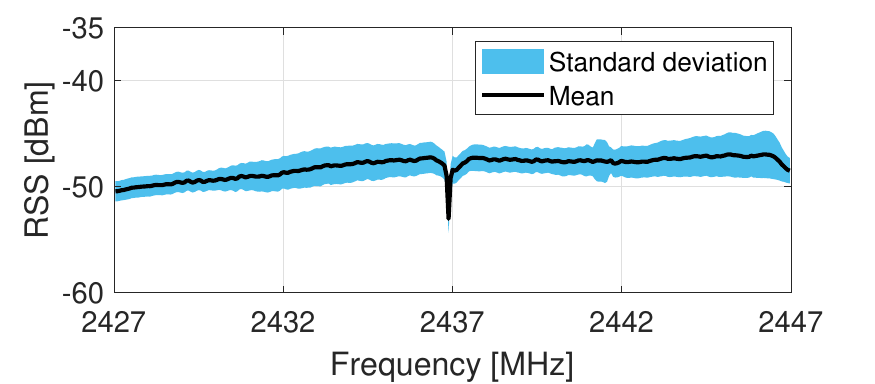}}
    
    \subfigure[RSS versus altitude: Channel 11.]{\includegraphics[trim={0.1cm, 0, 1.3cm, 0.3cm},clip, width=0.29\textwidth]{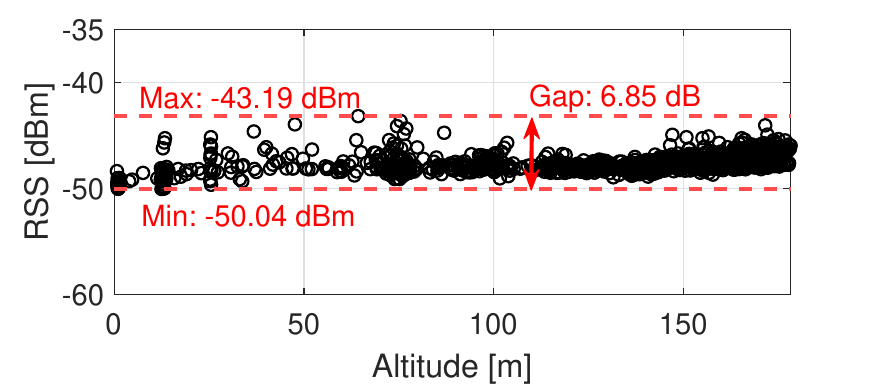}
    \label{fig:rss_altitude_2024_Packapalooza_WiFi_channel11}}
    \subfigure[RSS versus time with altitude: Channel 11.]{\includegraphics[trim={0.1cm, 0, 0cm, 0cm},clip, width=0.32\textwidth]{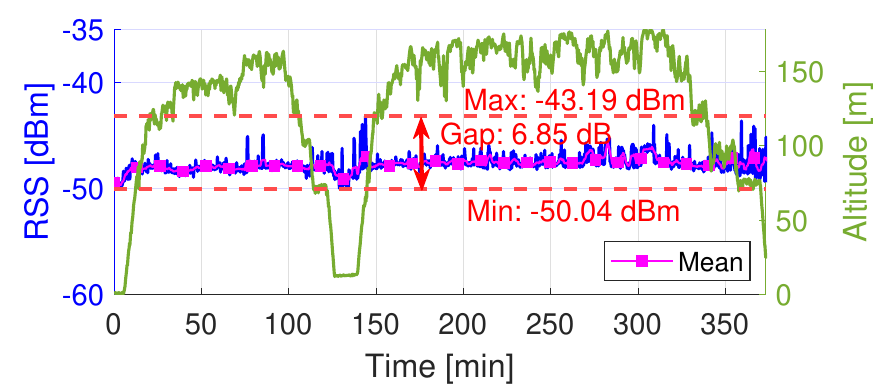}
    \label{fig:rss_time_altitude_2024_Packapalooza_WiFi_channel11}}
    \subfigure[RSS versus frequency: Channel 11.]{\includegraphics[trim={0.1cm, 0, 0.8cm, 0.3cm},clip, width=0.29\textwidth]{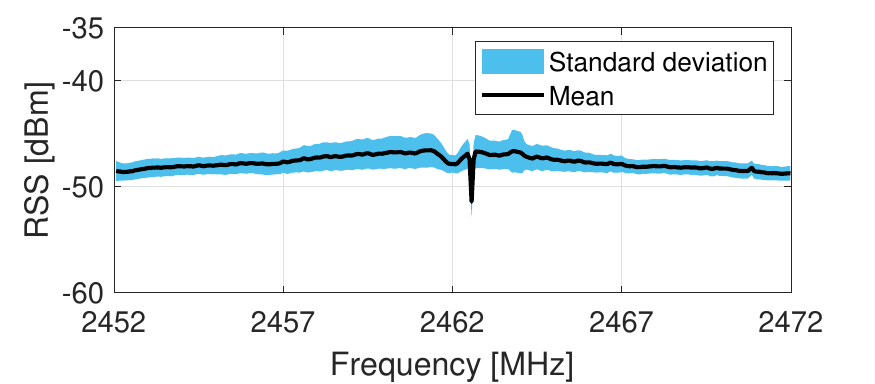}}%\vspace{-0.2cm}
    \caption{RSS measurements of non-overlapping channels in $2.4$~GHz ISM band over the urban areas of the 2024 Packapalooza Festival.}%\vspace{-0.4cm}
    \label{fig:rss_altitude_freq_2024_Packapalooza_WiFi}
\end{figure*}

A helikite equipped with a software-defined radio (SDR) and a GPS receiver was floated without an external power source. The 3D trajectory and altitude over time are shown in Fig.~\ref{fig:experiement_setups_helikite}. By capturing variations in the received signal strength (RSS) in the specific bands of interest, which act as interference sources in our UL asymmetric model, the measurement campaign provides physical evidence of interference in RC-UAV links. The SDR module keeps sweeping the range of spectrum from $87$~MHz to $6$~GHz to capture the RSS. The sampling rate and central frequency shift of the sweep are $30.72$~MHz and $25.68$~MHz, respectively. After the measurement campaign, we extracted the UL band of LTE and 5G, and $2.4$~GHz band of the ISM bands measurements, where the information of the bands of our interest is summarized in Table~\ref{tab:band_information_table}. The measured datasets are publicly available at~\cite{aerpawWebsite}.

\subsection{Helikite Measurement Campaign Results}\label{ch:helikite_results}
The RSS measurements obtained from the helikite in the urban areas during the $2024$ Packapalooza Festival are demonstrated in Figs.~\ref{fig:rss_altitude_freq_2024_Packapalooza_LTE} through~\ref{fig:rss_altitude_freq_2024_Packapalooza_WiFi} for LTE, 5G, and Wi-Fi, respectively. The measurement results have the following structure for each band of interest: 1) RSS versus altitude; 2) RSS versus time along with altitude variation, where blue curves correspond to the RSS (left y-axis), green curves show the altitude (right y-axis), and magenta curves show the moving averaged RSS with the duration of approximately $2.5$ minutes; and 3) RSS versus frequency.

\renewcommand{\arraystretch}{1.2}  
\renewcommand{\multirowsetup}{\centering}  
\begin{table}[t!]\vspace{0.05in}
    \centering
    \caption{Summary of LTE, 5G, and Wi-Fi bands analyzed in the helikite measurement campaign (modified from~\cite{Amir_table_paper}).}%\vspace{-0.2cm}
    \begin{tabular}{|c|c|c|c|c|}
    \hline
        \textbf{Standards} & \makecell{\textbf{Band or} \\ \textbf{channel No.}} & \makecell{\textbf{Duplex} \\ \textbf{mode}} & \makecell{\textbf{UL Band} \\ \textbf{[MHz]}} & \textbf{Operators} \\ \hline

        \multirow{4}{*}{LTE} 
          & $12$  & FDD & $698$-$716$ & \makecell{AT\&T \\ Verizon \\ T-Mobile } \\ \cline{2-5}
          & $13$ & FDD & $777$-$787$ & Verizon \\ \cline{2-5}
          & $14$ & FDD & $788$-$798$ & \makecell{AT\&T \\ FirstNet} \\ \cline{2-5}
          & $41$ & TDD & $2496$-$2690$ & T-Mobile  \\ \hline

        \multirow{3}{*}{\makecell{\\5G}} 
         & n$5$  & FDD & $824$-$849$ &  \makecell{AT\&T \\ Verizon} \\ \cline{2-5}
         & n$71$  & FDD & $663$-$698$ & T-Mobile \\ \cline{2-5}
         & n$77$  & TDD & $3700$-$3980$ & \makecell{AT\&T \\ Verizon \\ T-Mobile}  \\ \hline

        \multirow{3}{*}{\makecell{Wi-Fi \\ ($2.4$~GHz \\band)} }
            & $1$  & TDD & $2412$-$2432$ & \\ \cline{2-4}
            & $6$ & TDD & $2437$-$2457$ & ISM band \\ \cline{2-4}
            & $11$ & TDD & $2462$-$2482$ & \\ \hline
    \end{tabular}%\vspace{-0.4cm}
    \label{tab:band_information_table}
\end{table}

\subsubsection{LTE Bands}

In Fig.~\ref{fig:rss_altitude_freq_2024_Packapalooza_LTE}, the RSS measurements of LTE bands are shown. An altitude-dependent RSS pattern is observed in Bands $12$ and $41$, as demonstrated in Figs.~\ref{fig:rss_altitude_2024_Packapalooza_LTE_band12} and~\ref{fig:rss_altitude_2024_Packapalooza_LTE_band41}. In both bands, the stronger RSS tends to appear at higher altitudes due to the high chance of LoS signal reception. The gaps between the maximum and minimum RSS exceed $20$~dB and almost reach around $30$~dB for bands $12$ and $41$, respectively. The altitude-dependent variation is consistently observed when the altitude of the helikite is varied, e.g., the time duration under $20$~minutes and between $120$ to $140$~minutes in Figs.~\ref{fig:rss_time_altitude_2024_Packapalooza_LTE_band12} and~\ref{fig:rss_time_altitude_2024_Packapalooza_LTE_band41}. In the frequency domain, the RSS is distributed in a piecewise subdivided pattern over the allocated bandwidth. The band $12$ is subdivided into four parts, and the band $41$ is subdivided into two parts, as shown in Figs.~\ref{fig:rss_frequency_2024_Packapalooza_LTE_band12} and~\ref{fig:rss_frequency_2024_Packapalooza_LTE_band41}, respectively. These subdivisions may reflect the resource allocation or scheduling mechanisms used by the service provider.

In contrast, in the bands $13$ and $14$, the RSS has a relatively steadier pattern than the other bands, which have gaps between the maximum and minimum of RSS as $9.22$~dB and $5.95$~dB for bands $13$ and $14$, respectively. A similar flat pattern is also observed in the frequency domain, which shows almost flat and uniform over the whole allocated spectrum.

\subsubsection{5G Bands}

The RSS measurements of 5G bands are shown in Fig.~\ref{fig:rss_altitude_freq_2024_Packapalooza_5G}. A similar altitude-dependent RSS pattern is observed with the band n77 in Fig.~\ref{fig:rss_altitude_2024_Packapalooza_5G_bandn77} with a gap between the maximum and minimum of RSS as $28.25$~dB. On the other hand, the bands n5 and n71 exhibit a relatively steady pattern with RSS gaps of $9.95$~dB and $12.92$~dB, respectively. Moreover, similar to LTE bands $12$ and $41$, the 5G band n77 shows noticeable altitude-dependent variations during the periods under $20$~minutes and between $120$ to $140$~minutes, as shown in Fig.~\ref{fig:rss_time_altitude_2024_Packapalooza_5G_bandn77}. In the frequency domain, the subdivided pattern, previously seen in LTE, is also observed in 5G bands. Specifically, the bands n71 and n77 show three and four parts, respectively, while band n5 shows a uniform pattern across the spectrum.

\subsubsection{ISM Bands}

Fig.~\ref{fig:rss_altitude_freq_2024_Packapalooza_WiFi} presents the RSS measurements of $2.4$~GHz ISM bands. It is observed that channel 1 shows the largest RSS gap of $16.66$~dB among other channels, followed by $8.47$~dB of channel 6 and $6.85$~dB of channel 11. Identical to LTE and 5G, the RSS tends to be higher at higher altitudes, as shown in Fig.~\ref{fig:rss_altitude_2024_Packapalooza_WiFi_channel1}. Notably, a higher range of variation of RSS is observed at the second phase of ascending (time duration after $170$~minutes), as shown in Fig.~\ref{fig:rss_time_altitude_2024_Packapalooza_WiFi_channel1}, compared to the first ascending phase of helikite (from $0$ to $120$~minutes). This larger range of variation in the later phase (after $170$~minutes) is attributed to the higher density of people in the festival area during that period.

\enlargethispage{-0.05in}

In the frequency domain, channel 1 also shows the largest standard deviation range among the other channels. Channels 6 and 11 demonstrate a relatively steady pattern compared to channel 1. Additionally, in all channels, sharp drops in the center of each spectrum can be observed, which is caused by the null subcarrier from the orthogonal frequency division multiplexing (OFDM) process.

\section{System Model}\label{CH:sys}

\subsection{A2G Network Model}
In our A2G network model, a UAV functions as the cellular base station, and the RC on the ground serves the UE, as shown in Fig.~\ref{fig:A2G_model}. As seen in Section~\ref{ch:helikite_results}, the RSS measurements at the helikite, which are considered as interference in the UL, tend to increase as the altitude of the UAV increases due to the higher probability of LoS interference sources~\cite{Amir_table_paper}. Thus, the UAV experiences a lower SINR in the UL, resulting in an UL/DL SINR asymmetry caused by the LoS interference sources on the ground. This lower UL SINR is expected to degrade HARQ indicator feedback, bringing throughput degradation. 

\subsection{UL/DL SINR Asymmetry}\label{Sec:asymmetry}
The PUCCH symbol carrying the HARQ indicator feedback at the receiver side is generated and transmitted to the transmitter. The structure and procedure for generating the PUCCH symbol in LTE and 5G are summarized in~\cite{journal_work_arxiv}. To investigate the impact of SINR asymmetry in UL and DL, we assume the SINR for UL is worse than DL due to LoS interference at the UAV. The UL SINR is modeled as a simple bias applied to the DL SINR, which can be expressed as
\begin{equation}
    \gamma_{\mathrm{UL}}=\gamma_{\mathrm{DL}} - \delta,
\end{equation}
where $\gamma_{\mathrm{UL}}$ and $\gamma_{\mathrm{DL}}$ denote the UL and DL SINRs in dB, respectively, and $\delta$ represents the asymmetry bias in dB, facilitating UL SINR asymmetry.  

\section{Numerical Analysis}\label{ch:numerical_analysis}
\subsection{Performance Evaluation Setups}

\begin{figure}[t!]\vspace{0.04in}
    \centering
    \includegraphics[width=0.97\columnwidth]{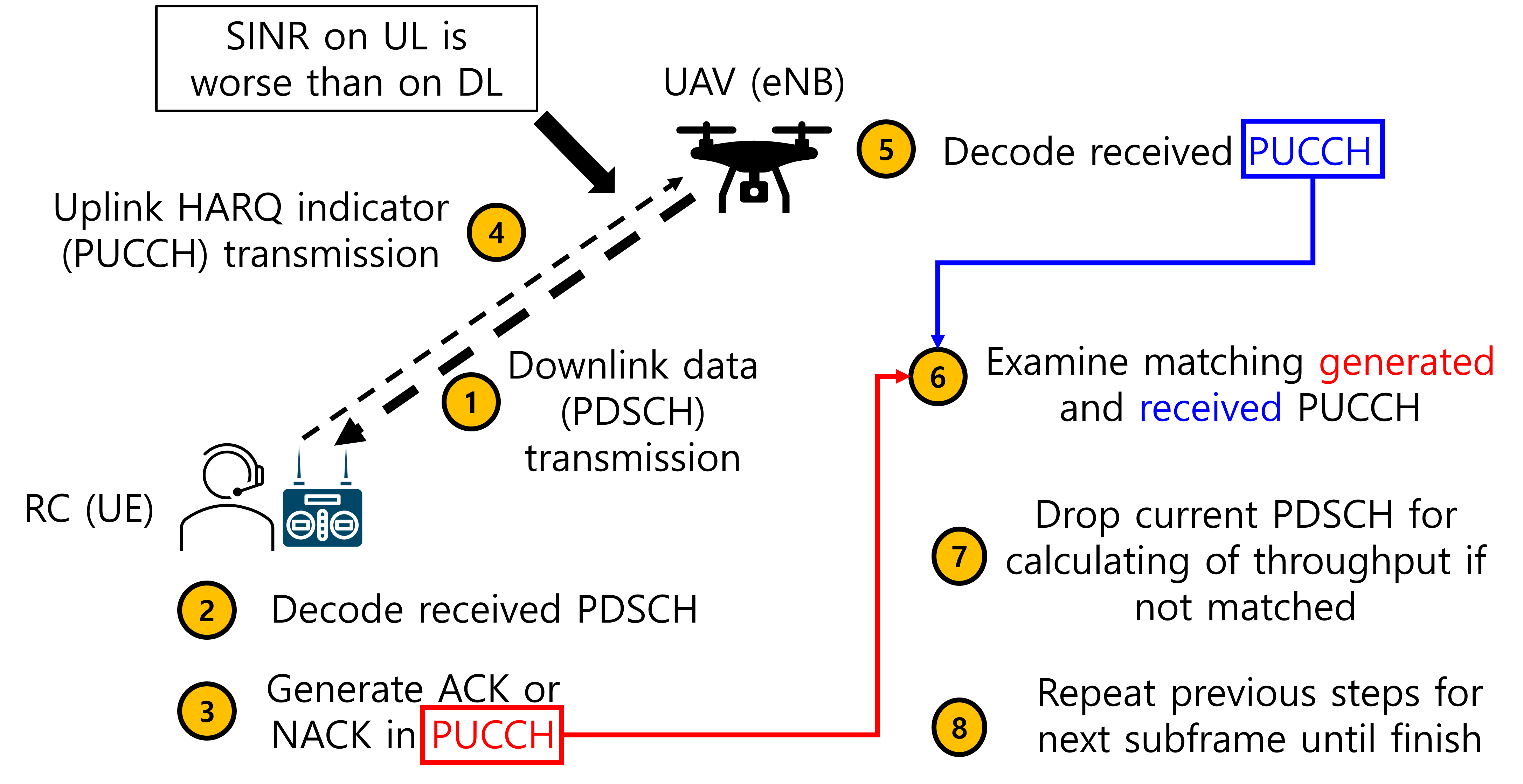}%\vspace{-0.2cm}
    \caption{UL/DL SINR asymmetry analysis procedure.}%\vspace{-0.4cm}
    \label{fig:uplink_asymmetry_procedure}
\end{figure}

We consider a point-to-point communication scenario between the UAV and the remote controller on the ground. We also adopt a Rayleigh fading channel with the extended vehicular A model (EVA) and maximum Doppler frequency of $5$~Hz. MATLAB's LTE and 5G Toolboxes~\cite{matlab_LTE, matlab_5G} are used for carrying out all the simulations. PUCCH format 1a, which carries one bit of HARQ indicator, is adopted for LTE simulation. For the 5G, PUCCH format 1 with proper parameters, e.g., hopping options and number of resource blocks, is selected for direct comparison with LTE.

The procedure for throughput evaluation under UL/DL asymmetry is illustrated in Fig.~\ref{fig:uplink_asymmetry_procedure}. First, the UAV transmits the DL data via the PDSCH. The RC unit decodes the received PDSCH subframe and sends a HARQ indicator feedback through PUCCH, which is then decoded at the UAV. Due to the asymmetric interference at the UL, the SINR of the UL is worse than that of the DL. To determine whether the feedback was correctly received, the UAV compares the decoded PUCCH with the originally transmitted PUCCH from the RC. If they do not match, the PUCCH is considered lost, and the corresponding PDSCH subframe is excluded for throughput calculation. The throughput at a given SINR can be calculated as
\begin{equation}
    TH=D/(N_\mathrm{SF}\times 10^{-3}),
\end{equation}
where $D$ is the accumulated size of the correctly decoded, including PDSCH and PUCCH, transport block, and $N_{\mathrm{SF}}$ is the number of transmitted subframes. 

\begin{table}[t!]\vspace{0.04in}
    \centering    
    \caption{Simulation parameters for throughput evaluation.}%\vspace{-0.2cm}
    \begin{tabular}{|c|c|c|}
        \hline  \textbf{Parameter}   & \textbf{Value} & \textbf{Description} \\ \hline
       $N_{\rm SF}$  & $500$ & Maximum number of subframe index \\ \hline
       % $N_{\rm HARQ}$ & $8$ & Number of parallel HARQ streams \\ \hline
       % $N_{\rm tr,max}$ & $4$ & Maximum number of transmission \\ \hline
       % $N_{\rm RB}$ & $50$ & Number of resource blocks \\ \hline
       TM & TM1 (SISO) & Transmission mode \\ \hline
       Modulation & QPSK & Modulation scheme \\ \hline 
       $c$ & $0.5$ & Coding rate \\ \hline
       $W_{\mathrm{DL}}$ & $10$ MHz & DL Bandwidth ($50$ resource blocks) \\ \hline
       $W_{\mathrm{UL}}$ & $1.4$ MHz & UL Bandwidth ($6$ resource blocks) \\ \hline
       Duplex Mode & FDD & Frequency division duplexing \\ \hline       
    \end{tabular}%\vspace{-0.4cm}
    \label{tab:sim_params}
\end{table}

The following assumptions are made for throughput evaluation: 1) the maximum number of transmission subframes is set to 500; 2) transmission mode 1 of single-input single-output (SISO) antenna settings with $50$ and $6$ resource blocks are employed for DL and UL, respectively; 3) quadrature phase shift keying (QPSK) with coding rate of $0.5$ is adopted; and 4) all overheads in the resource grid are taken into account for data transmission and throughput evaluation. The key parameters for throughput evaluation simulation are shown in Table~\ref{tab:sim_params}. The MATLAB scripts for simulation are publicly available at~\cite{github_repo}.

\subsection{Numerical Results}\label{ch:results}
\begin{figure}[t!]\vspace{0.04in}
    \centering 
    \subfigure[LTE.]{    \includegraphics[trim={0.4cm, 0, 1.1cm, 0.35cm},clip, width=0.85\columnwidth]{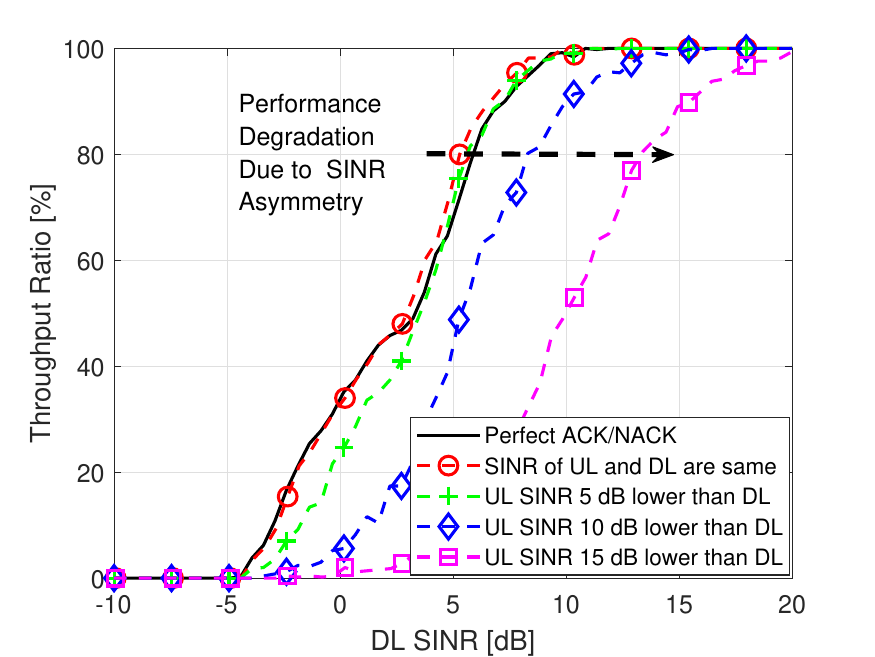}
    \label{fig:UL_Rayleigh_LTE}}
    \subfigure[5G.]{    \includegraphics[trim={0.4cm, 0, 1.1cm, 0.35cm},clip, width=0.85\columnwidth]{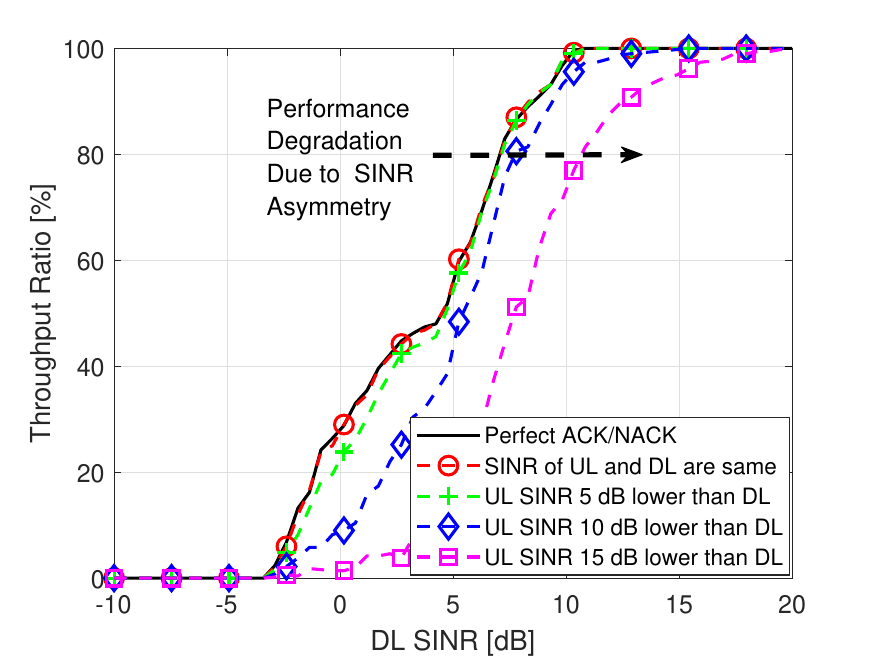}
    \label{fig:UL_Rayleigh_5G}}
    \caption{LTE and 5G DL throughput evaluation over the Rayleigh fading channel with UL/DL SINR asymmetry.}\vspace{-0.4cm}
    \label{fig:UL_Rayleigh_LTE_5G}
\end{figure}

The numerical results of the DL throughput evaluation under SINR asymmetry between UL and DL are demonstrated in Fig.~\ref{fig:UL_Rayleigh_LTE_5G}~\cite{journal_work_arxiv}. The x-axis represents the SINR in the DL channel, while the UL channel is set to one of four conditions: 1) identical SINR to DL, 2) UL SINR $5$~dB worse than the DL, 3) UL SINR $10$~dB worse than the DL, and 4) UL SINR $15$~dB worse than the DL. The perfect HARQ indicator reception cases are also included for comparison purposes. Since the size of the transport block in LTE and 5G with the given modulation and coding scheme is different, we introduce the throughput ratio as a normalized throughput metric. This can be expressed as $TH_{\mathrm{ratio}}=TH/TH_{\mathrm{max}}\times100$, where $TH$ denotes the throughput at a given SINR point and $TH_{\mathrm{max}}$ is the maximum achievable throughput with a given modulation and coding scheme.

The throughput performance with UL SINR identical is consistent in both LTE and 5G, while $5$~dB lower UL cases only show slight performance degradation. These indicate that the impact of the HARQ indicator loss in those settings is limited. However, noticeable throughput degradation can be observed as the UL asymmetry worsens, especially from $0$~dB to $15$~dB. This demonstrates that the impact of HARQ indicator decoding error from the asymmetric UL degrades the throughput performance. It is also observed that 5G is more robust against this UL asymmetry than LTE. Moreover, 5G demonstrates greater robustness to UL asymmetry than LTE, as confirmed by PUCCH block error rate evaluation in Fig.~15 of~\cite{journal_work_arxiv}. Based on both the measurement and simulation results in Sections~\ref{CH:measurement} and~\ref{ch:results}, it is essential to maintain a balanced UL and DL to fully exploit the achievable throughput capacity in the UAV to RC links.

\vspace*{0.02in}
\section{Conclusions}
In this paper, considering UL and DL SINR asymmetry, we evaluated the throughput performance of A2G links. To obtain physical evidence of UL asymmetry, we conducted the measurement campaign using a helikite platform at the Main Campus area of NC State University during the 2024 Packapalooza festival. The altitude-dependent pattern in which the RSS increased as the altitude of the helikite increased was observed due to the higher chances of LoS interference. To evaluate the impact of the UL asymmetry, we conducted MATLAB simulations using LTE and 5G Toolboxes. Substantial throughput degradation, demonstrating the impact of UL asymmetry in HARQ indicator feedback, was confirmed by our numerical results. Consequently, our measurement and simulation results highlighted the importance of balanced UL and DL for UAVs and RC links to achieve the maximum capacity of the given link.
% \section*{Acknowledgment}

\bibliographystyle{IEEEtran}
\bibliography{ref}
\typeout{CHECK: columnsep=\the\columnsep}
\end{document}